\documentclass[11pt]{article}
\usepackage{moriond,epsfig}

\def\Journal#1#2#3#4{{#1} {\bf #2}, #3 (#4)}

\def\be{\begin{equation}}
\def\ee{\end{equation}}
\def\bea{\begin{eqnarray}}
\def\eea{\end{eqnarray}}

\def\beq#1{\begin{equation}\label{#1}}
\def\eeq{\end{equation}}
\def\beqa#1{\begin{eqnarray}\label{#1}}
\def\eeqa{\end{eqnarray}}

\def\comment#1{\relax}

\begin{document}
\vspace*{4cm}
\title{STELLAR BLACK HOLE MASS FUNCTION: DETERMINATION
AND POSSIBLE IMPLICATIONS FOR FUNDAMENTAL GRAVITY}

\author{ K.A. POSTNOV }

\address{Sternberg Astronomical Institute, 
13 Universitetskij prosp., \\
Moscow 119992, Russia}

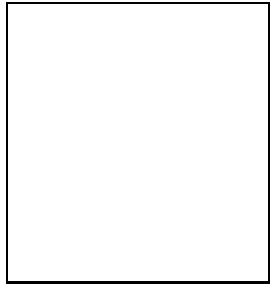
\maketitle\abstracts{We discuss masses of stellar black holes
found in binary systems and errors in their determination.
The observed mass distribution has a broad shape within the range 
$4-16 M_\odot$ without visible concentration to some 
preferred value. On the other hand, the initial black hole
mass funciton as inferred from observations of luminous
X-ray sources in other galaxies shows a power-law form.
We show that both dynamically obtained black hole mass function
and derived from X-ray observations can be made consistent
in the frame of the hypothesis of enhanced black hole evaporation
on a RSII brane for reasonable values of the warp-factor of
the Anti-de-Sitter bulk space.}

\section{Introduction}

There is a growing number of dynamical mass determinations of 
invisible compact objects in binary systems that are believed to be
black holes (BH) primarily for their large mass
($> 3M_\odot$) and lack of evidence for
having solid surface (see e.g. Orosz~\cite{Orosz_03}, 
Cherepashchuk~\cite{Cher_03} for recent reviews).
The mass distribution of compact objects in binary systems 
shows that 
masses of neutron stars (NS) are concentrated within 
a narrow range $M_{NS}=(1-2) M_\odot$ while all stellar 
BH masses fall within a wide interval $\sim 4-16 M_\odot$
with a ``mass gap'' 2-4 $M_\odot$ 
(e.g. Bailyn et al.~\cite{Baylin_ea98},~\cite{Cher_01} 
and references therein). Indirect information on BH masses
in binary systems can also be obtained from examining 
properties of X-ray luminosity function of the most luminous
(with X-ray luminosity $L_x>10^{39}$~erg/s)  
X-ray point sources observed in external galaxies by 
X-ray satellites assuming these sources to contain stellar
mass BH and their luminosities being at the Eddington 
level. While dynamical BH masses appear to have a 
nearly flat distribution within the 4-16 $M_\odot$ range
$dN/dM\sim M^0$, 
the X-ray luminosity function evidences for a power-law 
BH mass distribution $dN/dM\sim M^{-1.9...-2.2}$.
Unless some selection effects shape the form of
BH mass distribution in both cases, BH mass functions derived
from dynamical and X-ray observations are very different. 
Here we hypothesize that the above properties of the observed
BH mass function can be explained assuming an initial
power-law BH mass function $(dN/dM)_0\sim M^{-\alpha}$
with $\alpha \sim -2...-3$ and secular mass decrease of BH mass
which is possible in the framework of some modern 
theories of multidimensional gravity.  
  
\section{Determination of black hole masses and their errors}

The detailed analysis of systematic errors in 
determination of BH masses from observations 
which can affect the observed BH mass distribution 
is given elsewhere~\cite{PostnovCherepashchuk_03} and
we only briefly describe them here. 

\subsection{Dynamical BH mass determination in binary systems}

The basic information 
on the mass of the invisible component in a binary system is 
obtained from the mass function of the optical companion 
\beq{mass_f}
f_v(m)=\frac{m_x^3\sin^3i}{(m_x+m_v)^2}=1.038\times 10^{-7}
K_v^3P(1-e^2)^{3/2}\,,
\eeq
where $K_v$ is the semi-amplitude of the optical
star radial velocity curve (in km/s), $m_v$ and $m_x$ 
are masses of the optical and invisible star
(in solar units), respectively, $P$ is the orbital period
of the binary system (in days), $e$ is the orbital 
eccentricity and $i$ is the binary inclination angle 
with respect to the line of sight. The mass of the invisible companion 
as derived from the optical mass function is
\beq{m_x}
m_x=f_v(m)\left(1+\frac{1}{q}\right)^2\frac{1}{\sin^3 i}\,.
\eeq 
where $q=m_x/m_o$ is the mass ratio. 
The only value of $f_v(m)$  immediately 
provides the lower limit $m_x \ge f_v(m)$. 
More precise value of $m_x$ requires the knowledge
of $q$ and $\sin i$. 
 
a) Optical star form effects.

If the mass ratio $q<1$ (high-mass X-ray binaries, HMXB, e.g. Cyg X-1, 
LMC X-1, SS 433)), the system's barycenter lies within the
optical star body and tidal distortions of the optical star form  
mostly influence the spectral line profiles which are used to determine
the radial velocity curve~\cite{Cher_96}. Another distortion 
of radial velocity curve and its semi-amplitude $K_v$  in 
HMXB is due to variable selective absorption of light by 
strong stellar wind of the massive O-B optical star~\cite{Milgrom_78}.

In low-mass X-ray binaries (LMXB) $q>1$ the system's barycenter
is outside the optical star so the effects of the star form can be
neglected. Stellar wind effects from low-massive A-M dwarfs in these
systems are also insignificant.     

b) Stellar rotation effects.

The mass ratio $q$ is usually derived from rotational
broadening of absorption lines 
in the optical star spectrum. Indeed, assuming synchronous 
axial and orbital rotation, one can find~\cite{Orosz_03}
\beq{q}
v\sin i = 0.462 K_v q^{-1/3}\left(1+\frac{1}{q}\right)^{2/3}\,.
\eeq 
The determination of $v\sin i$ from usual analysis of 
absorption line profiles can not be made better than to within the 10-20\% 
errors due to X-ray heating of the optical star atmosphere
~\cite{Antokhina_ea03}. However, parameters $q, i$ can be
derived from orbital variability of the absorption 
line profiles using new method proposed by
Antokhina and Cherepashchuk~\cite{AntokhinaCher_97} and
Shahbaz~\cite{Shahbaz_98}. This allows the strong reducing of  
errors in the parameter $q$ determination. 

c) Binary inclination angle effects.

Uncertainties in the binary inclination angle 
$i$ provide the largest error in the mass determination. 
The usual way of measuring $i$ is from the ellipticity 
effect of the optical star~\cite{Lyuty_ea73}. 
The main uncertainty here comes from model-dependent contributions
of other emitting structures 
(gaseous stream, accretion disk) 
into the total  optical or infrared 
variability of the system and is especially important ($>50\%$) 
in LMXB (in which most BH candidates have been discovered).
The new method of determination of  parameters
$q, i$ based on the analysis of the
orbital variation of the absorption line profiles
~\cite{AntokhinaCher_97},\cite{Shahbaz_98} is independent
of contribution from other gas structures in the binary system.
However, it can be applied  
at a very high spectral resolution of 50000-100000
and can be realized only on very large telescopes.  
     
So, with the 90\% certainty we can state that 
masses of stellar BH measured dynamically in binary systems
span a wide rage 4-16 $M_\odot$ without visible concentration
to some value, i.e. $dN/dM\sim M^\alpha$ with $\alpha\sim 0$. 

\subsection{X-ray luminosity function of ultra-luminous
X-ray sources}

Ultra-luminous X-ray sources (ULX) are point-like sources with 
persistent X-ray 
luminosity $L_x> 2-4 10^{38}$ erg/s corresponding to a 1.4 $M_\odot$ 
neutron star. They have been   
discovered by several X-ray satellites (mostly by {\it Chandra}, 
see e.g. \cite{Humphrey_ea03}
and references therein) in other galaxies. Although some of them
proved to be another galaxies seen through the observed
galaxy disk (see \cite{Masetti_ea03} for a recent identification), 
they constitute the most luminous population of X-ray sources in 
galaxies and 
based on their X-ray spectral fitting
can be identified with accreting stellar-mass BH in binary systems 
\cite{TerashimaWilson_03}. 

A general analysis of X-ray source populations in 
nearby galaxies was recently carried out by Grimm et al. 
\cite{Grimm_ea02}. It revealed the universal power-law shape 
of the average X-ray luminosity function $dN/dL_x\sim L_x^{-1.6}$ in 
a wide range of X-ray luminosities $10^{34}-10^{39}$ erg/s 
with a steeper decline $\propto L_x^{-2.2}$ for the most luminous sources 
$L_x>2\times 10^{39}$ erg/s. 
The universal shape of the X-ray luminosity function below
$(2-4)\times  10^{38}$ erg/s can be explained on the very general
grounds by accretion in binary systems~\cite{Postnov_03}.  
The cut-off in the luminosity
function at $L_x>10^{39}$ erg/s (i.e. 
for ultra-luminous X-ray sources) 
appears to be a general feature of X-ray source
populations in galaxies (see e.g. a recent analysis of {\it Chandra} 
observations of NGC 5194 in \cite{TerashimaWilson_03})
reflecting the presence of binaries with BH accreting at 
the Eddington limit. This allows us to assume that for BH in 
these systems 
$dN/dM=(dN/dL_x)(dL_x/dM) \sim L_x^{-1.9...-2.2}\sim M^{-1.9...-2.2}$.    

Uncertainties in this method of determining BH mass distribution 
are due to possible
beaming of X-ray emission from binary systems (the value of 
the X-ray luminosity is usually obtained from the observed X-ray fluxes
assuming spherical symmetry). At present, this uncertainty is hard to 
estimate, but similar shapes of X-ray luminosity functions of 
point-like sources 
in different galaxies and the very possibility of obtaining
a universal power-law form for $dN/dL_x$ which 
is simply scaled by the star formation rate from galaxy to galaxy 
(as argued in \cite{Grimm_ea02}) 
seem encouraging.  

\section{Black hole mass function and enhanced black hole evaporation}

Clearly, the decreasing power-law 
shape of BH mass distribution is drastically different
from the nearly flat form derived from dynamical BH 
mass determinations. 
First, we should note that if ULX are actually 
BH accreting at the Eddington limit, the mass
accretion rates $\dot M$ in these binaries
must be above $\sim 10^{-7} M_\odot$/year, i.e.
such systems must be HMXB with ages $<10^7$ years. In contrast, 
most BH with dynamically determined masses reside in
transient LMXB (X-ray novae)\cite{Orosz_03}, \cite{Cher_03}
which are old systems with ages $>10^8$ years. 
 
Let us consider the  distribution function $f(M)\equiv dN/dM$ of 
a population of sources with changing mass. Let the initial
mass distribution be $f_0(M)$ within 
the mass range $[M_{\min}, M_{max}]$ 
and the law of the mass change 
be $M(t)$. In the stationary case the evolution of mass distribution 
is described by the one-dimensional kinetic equation 
\beq{kin}
\frac{\partial }{\partial M}\left[f(M) \dot M \right] = f_0(M)
\eeq 
so the stationary distribution function for $\dot M>0$ will be  
\beq{f_st}
f(M)=\frac{\int_{M_{min}}^{M}f_0(M')dM'}{\dot M}\,, \quad M<M_{max}
\eeq
while for $M\ge M_{max}$ the stationary distribution is independent of
the initial mass function and is determined only by 
the mass change $\dot M$:
\beq{}
f(M)=\frac{\int_{M_{min}}^{M_{max}}f_0(M')dM'}{\dot M}=
\frac{\hbox{const}}{\dot M}\,, \quad M\ge M_{max}
\eeq
For example, assuming the initial power-law mass function 
$f_0(M)\propto M^{-\alpha_i}$ yields for $\alpha_i>1$  
$f(M)\sim [1-(M/M_{min})^{-\alpha_i+1}]/\dot M\sim 1/\dot M$ at $M>$ 
a few $M_{min}$, and for $\alpha_i <1$ 
$f_0(M)\sim [(M/M_{min})^{-\alpha_i+1}-1]/\dot M$. 
Since BH mass in accreting binaries can never increase 
faster than $\dot M\propto M$, the stationary distribution 
never (for any $\alpha_i$) 
decreases steeper than $1/M$. So the observed steep BH mass cut-off
(propto $M^{-1.9...-2.2}$) in ULX apparently evidences that 
stationarity arguments for these systems are inapplicable 
due to short time of Eddington-limited accretion in these systems
(typically $\sim 10^{5}$ years $\ll M/\dot M$) so that 
BH masses cannot change significantly. Then we stay with the
possibility that the observed steep BH mass function
reflects the initial BH mass distribution in these binaries.

In LMXB (in which most BH candidates are found) or 
for single BH (only two of them are suspected by 
gravitational microlensing experiments \cite{Bennett_ea02})
the situation can be quite different. Small mean 
accretion rates $\dot M \sim 10^{-9} M_\odot$/yr
typical for these systems are insignificant
for BH mass growth. Applying the same arguments
as above, we could expect the initial BH mass function to 
have a nearly flat shape in this case. However, there is
another possibility.

Let us consider the hypothesis of enhanced BH mass evaporation recently put
forward by Tanaka~\cite{Tanaka_02} and Emparan et al.~\cite{Emparan_ea02}.
These authors consider the RSII brane world models~\cite{RandalSundrum_99}
in which our Universe is localized on a 4D-brane embedded 
in an Anti-de-Sitter bulk characterized by the warp factor (curvature
length) $\cal L$. 
Such models admit macroscopic values of $\cal L$ up to 
0.1 mm so as not to contradict the existing laboratory
measurements \cite{LongPrice_03}. In this setup, the authors
~\cite{Tanaka_02},~\cite{Emparan_ea02} speculate that 
no stationary macroscopic BH can exist on the brane 
due to enhanced evaporation into (virtually unobservable)
CFT-modes (KK-gravitons). This process has not yet been calculated
properly so this hypothesis remains highly speculative (see e.g. 
Casadio 2003~\cite{Cassadio_03} who obtained different results). 
Nevertheless, we try to see what happens should the  
macroscopic BH evaporation actually exist.    

In this hypothesis, the evaporation time $\tau$ of a macroscopic BH
localized on the brane 
(i.e. with the horizon size $> {\cal L}$) is shorter than 
the classical Hawking evaporation time 
$\sim t_{Pl}(M/m_{Pl})^3$ by a huge factor $O({\cal L}/l_{Pl})^2$
(subscript {\it Pl} stands for the Planck units) 
\beq{}
\tau \simeq 10^2 [\hbox{yrs}] \left(\frac{M}{M_\odot}\right)^3 
\left(\frac{1 [\hbox{mm}]}{\mathcal L}\right)^2
\eeq
So the mass of an isolated BH decreases as $\dot M \sim M^{-2}$. 
It is easy to show that for $\cal L\sim 10^{-3}-10^{-4}$ mm 
the BH evaporation rate exceeds the mean accretion rate 
in LMXB for $M\sim 10 M_\odot$. So the present bounds 
$\cal L<0.1$~mm do not contradict this assumption.

Now for old BH in low-mass X-ray binaries 
we can find the stationary
shape of BH mass distribution function. For evaporation $\dot M<0$ so
the integration of Eq. (\ref{kin}) yields
\beq{f_st1}
f(M)=\frac{\int_M^{M_{max}}f_0(M')dM'}{\dot M}\,, \quad M>M_{min}
\eeq
and 
\beq{}
f(M)=\frac{\int_{M_{min}}^{M_{max}}f_0(M')dM'}{\dot M}=
\frac{\hbox{const}}{\dot M}\,, \quad M\le M_{min}
\eeq
Assuming as above $f_0(M)\sim M^{-\alpha_i}$ with $\alpha_i<1$ 
we have $f(M)\sim M^{-\alpha_i+3}$ for $M>M_{min}$ and 
$f(M)\sim M^2$ for $M\le M_{\min}$ (we always assume $M\ll M_{max}$).  
This is illustrated in Fig. 1. Clearly, if $\alpha_i\sim 2$, 
as can be inferred from ULX observations, the stationary 
BH mass distribution could be made much flatter by BH
evaporation. In addition, the steep
decrease of the stationary BH mass function ($f(M)\sim M^2$) 
below $M_{min}$ can be identified with apparent deficit 
of BH with masses below 4 $M_\odot$.

\begin{figure}
\centerline{\psfig{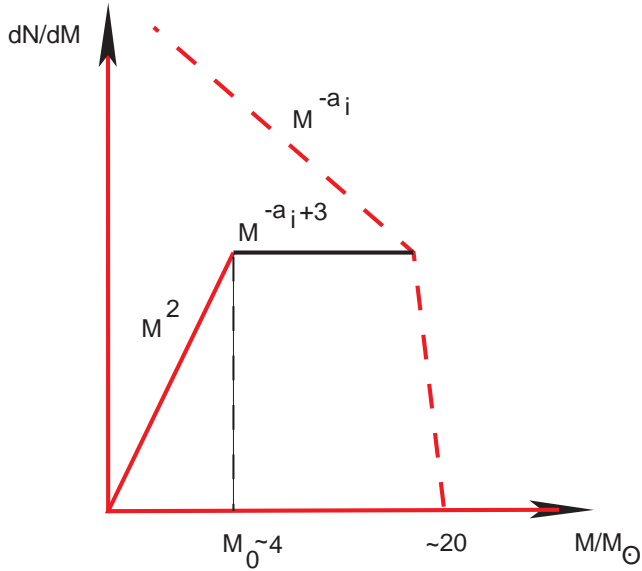}}
\caption{Schematic view of the expected stationary 
BH mass function (the solid curve) with the initial
power-law form 
$(dN/dM)_i\sim M^{-{a_i}}$ (the dashed line) 
obtained in the enhanced BH evaporation model on the RSII-brane.
The mass  $M_0$ corresponds to 
a minimal BH mass that can evaporate in the Hubble time.
}
\end{figure}

\section{Conclusions}

The steadily growing number of mass determinations of
stellar BH allows us to construct and study BH mass function.
It has apparently broad and near flat shape within the interval
$4-16 M_\odot$ without visible concentration at some value, which 
is opposite to what is observed for NS masses ($M_{NS}=1-2 M_\odot$). 
Possible systematic errors in BH mass determination, 
which mostly rely on the radial velocity curve of the optical 
companion, can be
reduced in future high-resolution spectroscopic observations.
    
BH mass function can also be inferred from observations of 
luminous X-ray binary systems (with $L_x>10^{39}$ erg/s) 
in other galaxies. The current 
{\it Chandra} observations suggest a steep decrease 
($dN/dL_x\sim L_x^{-2}$) in the 
X-ray luminosity function of point-like objects in galaxies.
If these sources are accreting BH at the Eddington limit, 
the initial BH mass function should have a similar slope
$f_0(M)\sim M^{-2}$, which disagrees with BH distribution 
found in galactic binary systems. 

We show here that in the framework of the enhanced
BH evaporation hypothesis \cite{Tanaka_02}\cite{Emparan_ea02} 
which can in principle take place 
in some multidimensional gravity models (of RSII type), the initial 
steep power-law BH mass function can be made flatter
to match the observed BH distribution in galactic binaries. 
This hypothesis also predicts low-mass BH 
(with $M$ less than a few solar masses) should be very 
rare. Discovery of such low-mass BH would strongly limit
this hypothesis. 
 
Concluding, we note that 
it is unclear at present how close to reality such multidimensional
gravity models are, so any test of their astrophysical 
predictions is very desirable. We hope that future 
precision measurements of BH masses (both single and in binaries) 
by traditional and new (e.g., gravitational 
waves from coalescing binaries with BH) astrophysical methods 
can be potentially 
very interesting in this respect. 
   
\section*{References}





\end{document}